\documentclass[12pt]{article}
\usepackage{a4wide}
\usepackage{latexsym}
\usepackage{amsmath}
\usepackage{amsfonts}
\usepackage{cite}
\usepackage{axodraw}

\usepackage{pslatex}
\usepackage[latin1]{inputenc}
\usepackage[T1]{fontenc}

\def\bq{\begin{eqnarray}}
\def\eq{\end{eqnarray}}
\def\eps{\varepsilon}

\begin{document}

\thispagestyle{empty}

\begin{flushright}
  MZ-TH/07-06
%  \\ version of \today
\end{flushright}

\vspace{1.5cm}

\begin{center}
  {\Large\bf Feynman integrals and multiple polylogarithms\\
  }
  \vspace{1cm}
  {\large Stefan Weinzierl\\
\vspace{2mm}
      {\small \em Institut f{\"u}r Physik, Universit{\"a}t Mainz,}\\
      {\small \em D - 55099 Mainz, Germany}\\
  } 
\end{center}

\vspace{2cm}

% abstract ---------------------------------------
\begin{abstract}\noindent
  {
In this talk I review the connections between Feynman integrals and multiple polylogarithms.
After an introductory section on loop integrals I discuss the
Mellin-Barnes transformation and shuffle algebras.
In a subsequent section multiple polylogarithms are introduced.
Finally, I discuss how certain Feynman integrals evaluate
to multiple polylogarithms.
   }
\end{abstract}

\vspace*{\fill}

% main text ------------------------------------
\newpage

\section{Introduction}
\label{sect:intro}

In this talk I will discuss 
techniques for the computation of loop integrals, which occur
in perturbative calculations in quantum field theory. 
Particle physics has become a field where precision measurements have become possible.
Of course, the increase in experimental precision has to be matched with more accurate calculations
from the theoretical side.
This is the ``raison d'\^etre'' for loop calculations: A higher accuracy is reached by including more terms
in the perturbative expansion.
The complexity of a calculation increases obviously with the number of loops, but also with the number of external particles
or the number of non-zero internal masses associated to propagators.
To give an idea of the state of the art, specific quantities which are just pure numbers have been computed
up to an impressive fourth or third order.
Examples are
the calculation of the 4-loop contribution to the 
QCD $\beta$-function \cite{vanRitbergen:1997va}, 
the calculation of the anomalous magnetic moment of the electron 
up to three loops \cite{Laporta:1996mq},
and the calculation of the ratio
of the total cross section for hadron production to the total
cross section for the production of a $\mu^+ \mu^-$ pair
in electron-positron annihilation to order $O\left( \alpha_s^3 \right)$ \cite{Gorishnii:1991vf}.
Quantities which depend on a single variable are known at the best to the third order. 
Outstanding examples are the computation
of the three-loop Altarelli-Parisi splitting functions 
\cite{Moch:2004pa,Vogt:2004mw}
or
the calculation of the two-loop amplitudes for the most interesting
$2 \rightarrow 2$ processes 
\cite{Bern:2000dn,Bern:2000ie,Bern:2001df,Bern:2001dg,Bern:2002tk,Anastasiou:2000kg,Anastasiou:2000ue,Anastasiou:2000mv,Anastasiou:2001sv,Glover:2001af,Binoth:2002xg}.
For the calculation of these amplitudes, the knowledge of certain highly non-trivial two-loop
integrals has been essential \cite{Smirnov:1999gc,Smirnov:1999wz,Tausk:1999vh}.
The complexity of a two-loop computation increases, if the result depends on more than one variable.
An example for a two-loop calculation whose result depends on two variables is the computation of the
two-loop amplitudes for $e^+ e^- \rightarrow \mbox{3 jets}$
\cite{Garland:2001tf,Garland:2002ak,Moch:2002hm}.
But in general, if more than one variable is involved, we have to content ourselves with next-to-leading order
calculations. An example for the state of the art is here the computation of the electro-weak corrections
to the process $e^+ e^- \rightarrow \mbox{4 fermions}$ \cite{Denner:2005es,Denner:2005fg}.

From a mathematical point of view loop calculations reveal interesting algebraic structures.
Multiple polylogarithms play an important role to express the results of loop calculations.
The mathematical aspects will be discussed in this talk.
Additional material related to loop calculations 
can found in the reviews \cite{Smirnov:2002kq,Grozin:2003ak,Weinzierl:2003jx,Weinzierl:2006qs}
and the book \cite{Smirnov:2004ym}.

This paper is organised as follows:
In the next section I review basic facts about Feynman integrals.
Section~\ref{sect:mellinbarnes} is devoted to the Mellin-Barnes transformation.
In section~\ref{sect:shuffle} algebraic structures like shuffle algebras are introduced.
Section~\ref{sect:polylog} deals with multiple polylogarithms.
Section~\ref{sect:laurent} combines the various aspects and shows, how certain Feynman integrals evaluate
to multiple polylogarithms.
Finally, section~\ref{sect:concl} contains a summary.

% ----------------------------------------------
\section{Feynman integrals}
\label{sect:feynman}

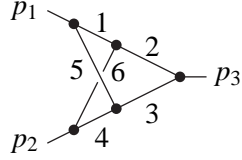
\begin{figure}
\begin{center}
\begin{picture}(57,40)(10,25)
\Line(50,30)(60,30)
\Vertex(50,30){2}
\Line(0,5)(50,30)
\Line(0,55)(50,30)
\Vertex(10,10){2}
\Vertex(10,50){2}
\Vertex(26,18){2}
\Vertex(26,42){2}
\Line(10,50)(26,18)
\Line(10,10)(19,28)
\Line(21,32)(26,42)
\Text(37,38)[bl]{\small$2$}
\Text(37,20)[tl]{\small$3$}
\Text(18,47)[bl]{\small$1$}
\Text(18,11)[tl]{\small$4$}
\Text(14,33)[r]{\small$5$}
\Text(24,32)[l]{\small$6$}
\Text(-3,55)[r]{\small $p_1$}
\Text(-3,5)[r]{\small $p_2$}
\Text(63,30)[l]{\small $p_3$}
\end{picture}
\end{center} 
\caption{\label{example_graph} An example of a two-loop Feynman graph with three external legs.}
\end{figure}

To set the scene let us consider a scalar Feynman graph $G$.
Fig.~\ref{example_graph} shows an example.
In this example there are three external lines and six internal lines. 
The momenta flowing in or out through the external lines
are labelled $p_1$, $p_2$ and $p_3$ and can be taken as fixed vectors.
They are constrained by momentum conservation: If all momenta are taken to flow outwards,
momentum conservation requires that
\bq
 p_1 + p_2 + p_3  & = & 0.
\eq
At each vertex of a graph we have again momentum conservation: The sum of all momenta flowing into the vertex
equals the sum of all momenta flowing out of the vertex.
A graph, where the external momenta determine uniquely all internal momenta is called a tree graph.
It can be shown that such a graph does not contain any closed circuit.

In contrast, graphs which do contain one or more closed circuits are called loop graphs.
If we have to specify besides the external momenta in addition 
$l$ internal momenta in order to determine uniquely all
internal momenta we say that the graph contains $l$ loops.
In this sense, a tree graph is a graph with zero loops and the graph in fig.~\ref{example_graph} contains two loops.
Let us agree that we label the $l$ additional internal momenta by $k_1$ to $k_l$.

Feynman rules allow us to translate a Feynman graph into a mathematical formula.
For a scalar graph we have substitute for each internal line $j$ a propagator
\bq
 \frac{i}{q_j^2-m_j^2+i\delta}.
\eq
Here, $q_j$ is the momentum flowing through line $j$. It is a linear combination of the external 
momenta $p$ and the loop momenta $k$:
\bq
 q_j & = & q_j(p,k).
\eq
$m_j$ is the mass of the particle of line $j$.
The propagator would have a pole for $p_j^2=m_j^2$, or phrased differently $E_j=\pm \sqrt{\vec{p}_j^2+m_j^2}$.
When integrating over $E$, the integration contour has to be deformed to avoid these two poles.
Causality dictates into which directions the contour has to be deformed. The pole on the negative real axis is avoided
by escaping into the lower complex half-plane, the pole at the positive real axis is avoided by a deformation 
into the upper complex half-plane. Feynman invented the trick to add a small imaginary part $i\delta$ to the 
denominator, which keeps track of the directions into which the contour has to be deformed.
In the following the $i\delta$-term is omitted in order to keep the notation compact.

The Feynman rules tell us also to integrate for each loop over the loop momentum:
\bq
 \int \frac{d^4k_r}{(2\pi)^4}
\eq
However, there is a complication:
If we proceed naively and write down for each loop an integral over 
four-dimensional Minkowski space, we end up with ill-defined integrals, since these integrals
may contain ultraviolet or infrared divergences!
Therefore the first step is to make these integrals well-defined by introducing a regulator.
There are several possibilities how this can be done, but the
method of dimensional regularisation 
\cite{'tHooft:1972fi,Bollini:1972ui,Cicuta:1972jf}
has almost become a standard, as the calculations in this regularisation
scheme turn out to be the simplest.
Within dimensional regularisation one replaces the four-dimensional integral over the loop momentum by an
$D$-dimensional integral, where $D$ is now an additional parameter, which can be a non-integer or
even a complex number.
We consider the result of the integration as a function of $D$ and we are interested in the behaviour of this 
function as $D$ approaches $4$.
It is common practice to parameterise the
deviation of $D$ from $4$ by
\bq 
 D & = & 4 - 2\eps.
\eq
The divergences in loop integrals will manifest themselves in poles in $1/\eps$. 
In an $l$-loop integral ultraviolet divergences will lead to poles $1/\eps^l$ at the worst, whereas
infrared divergences can lead to poles up to $1/\eps^{2l}$.
We will also encounter integrals, where the dimension is shifted by units of two.
In these cases we often write
\bq
\label{shifted_dim}
 D & = & 2m - 2\eps,
\eq
where $m$ is an integer, and we are again interested in the Laurent series in $\eps$.

Let us now consider a generic scalar $l$-loop integral $I_G$ 
in $D=2m-2\eps$ dimensions with $n$ propagators,
corresponding to a graph $G$.
Let us further make a slight generalisation: 
For each internal line $j$ the corresponding propagator
in the integrand can be raised to a power $\nu_j$.
Therefore the integral will depend also on the numbers $\nu_1$,...,$\nu_n$.
We define the Feynman integral by
\bq
\label{eq0}
I_G  & = &
 \left(  e^{\eps \gamma_E} \mu^{2\eps} \right)^l
 \int \prod\limits_{r=1}^{l} \frac{d^Dk_r}{i\pi^{\frac{D}{2}}}\;
 \prod\limits_{j=1}^{n} \frac{1}{(-q_j^2+m_j^2)^{\nu_j}}.
\eq
The momenta $q_j$ of 
the propagators are linear combinations of the external momenta and the loop
momenta.
In eq.~(\ref{eq0}) there are some overall factors, which I inserted for convenience:
$\mu$ is an arbitrary mass scale and the factor $\mu^{2\eps}$ ensures that the mass dimension of
eq.~(\ref{eq0}) is an integer.
The factor $e^{\eps \gamma_E}$ avoids a proliferation of Euler's constant
\bq 
 \gamma_E & = & \lim\limits_{n\rightarrow \infty} \left( \sum\limits_{j=1}^n \frac{1}{j} - \ln n \right)
 = 0.5772156649...
\eq
in the final result.
The integral measure is now $d^Dk/(i \pi^{D/2})$ instead of $d^Dk/(2 \pi)^D$, and each propagator
is multiplied by $i$.
The small imaginary parts $i\delta$ in the propagators are not written explicitly.

How to perform the $D$-dimensional loop integrals ?
The first step is to convert the products of propagators into a sum.
This can be done with the Feynman parameter technique.
In its full generality it is also applicable to cases, where each factor in the denominator is raised to 
some power $\nu$.
The formula reads:
\bq
 \prod\limits_{i=1}^{n} \frac{1}{P_{i}^{\nu_{i}}} 
 & = &
 \frac{\Gamma(\nu)}{\prod\limits_{i=1}^{n} \Gamma(\nu_{i})}
 \int\limits_{0}^{1} \left( \prod\limits_{i=1}^{n} dx_{i} \; x_{i}^{\nu_{i}-1} \right)
 \frac{\delta\left(1-\sum\limits_{i=1}^{n} x_{i}\right)}
      {\left( \sum\limits_{i=1}^{n} x_{i} P_{i} \right)^{\nu}},
 \;\;\;\;\;\;
 \nu = \sum\limits_{i=1}^{n} \nu_{i}. 
\eq
Applied to eq.~(\ref{eq0}) we have
\bq
 \sum\limits_{i=1}^{n} x_{i} P_{i} & = & \sum\limits_{i=1}^{n} x_{i} (-q_i^2+m_i^2).
\eq
One can now use translational invariance of the $D$-dimensional loop integrals and shift each loop
momentum $k_r$ to complete the square, such that the integrand depends only on $k_r^2$.
Then all $D$-dimensional loop integrals can be performed.
As the integrals over the Feynman parameters still remain,
this allows us to treat the
$D$-dimensional loop integrals for Feynman parameter integrals.
One arrives at the following Feynman parameter integral \cite{Itzykson:1980rh}:
\bq
\label{eq1}
I_G  & = &
 \left(  e^{\eps \gamma_E} \mu^{2\eps} \right)^l
 \frac{\Gamma(\nu-lD/2)}{\prod\limits_{j=1}^{n}\Gamma(\nu_j)}
 \int\limits_{0}^{1} \left( \prod\limits_{j=1}^{n}\,dx_j\,x_j^{\nu_j-1} \right)
 \delta(1-\sum_{i=1}^n x_i)\,\frac{{\mathcal U}^{\nu-(l+1) D/2}}
 {{\mathcal F}^{\nu-l D/2}}.
\eq
The functions ${\mathcal U}$ and $\mathcal F$ depend on the Feynman parameters.
If one expresses
\bq
 \sum\limits_{j=1}^{n} x_{j} (-q_j^2+m_j^2)
 & = & 
 - \sum\limits_{r=1}^{l} \sum\limits_{s=1}^{l} k_r M_{rs} k_s + \sum\limits_{r=1}^{l} 2 k_r \cdot Q_r - J,
\eq
where $M$ is a $l \times l$ matrix with scalar entries and $Q$ is a $l$-vector
with fourvectors as entries,
one obtains
\bq
 {\mathcal U} = \mbox{det}(M),
 & &
 {\mathcal F} = \mbox{det}(M) \left( - J + Q M^{-1} Q \right).
\eq
Alternatively,
the functions ${\mathcal U}$ and ${\mathcal F}$ can be derived 
from the topology of the corresponding Feynman graph $G$.
Cutting $l$ lines of a given connected $l$-loop graph such that it becomes a connected
tree graph $T$ defines a chord ${\mathcal C}(T,G)$ as being the set of lines 
not belonging to this tree. The Feynman parameters associated with each chord 
define a monomial of degree $l$. The set of all such trees (or 1-trees) 
is denoted by ${\mathcal T}_1$.  The 1-trees $T \in {\mathcal T}_1$ define 
${\mathcal U}$ as being the sum over all monomials corresponding 
to the chords ${\mathcal C}(T,G)$.
Cutting one more line of a 1-tree leads to two disconnected trees $(T_1,T_2)$, or a 2-tree.
${\mathcal T}_2$ is the set of all such  pairs.
The corresponding chords define  monomials of degree $l+1$. Each 2-tree of a graph
corresponds to a cut defined by cutting the lines which connected the two now disconnected trees
in the original graph. 
The square of the sum of momenta through the cut lines 
of one of the two disconnected trees $T_1$ or $T_2$
defines a Lorentz invariant
\bq
s_{T} & = & \left( \sum\limits_{j\in {\mathcal C}(T,G)} p_j \right)^2.
\eq   
The function ${\mathcal F}_0$ is the sum over all such monomials times 
minus the corresponding invariant. The function ${\mathcal F}$ is then given by ${\mathcal F}_0$ plus an additional piece
involving the internal masses $m_j$.
In summary, the functions ${\mathcal U}$ and ${\mathcal F}$ are obtained from the graph as follows:
\bq
\label{eq0def}	
 {\mathcal U} 
 & = & 
 \sum\limits_{T\in {\mathcal T}_1} \Bigl[\prod\limits_{j\in {\mathcal C}(T,G)}x_j\Bigr]\;,
 \nonumber\\
 {\mathcal F}_0 
 & = & 
 \sum\limits_{(T_1,T_2)\in {\mathcal T}_2}\;\Bigl[ \prod\limits_{j\in {\mathcal C}(T_1,G)} x_j \Bigr]\, (-s_{T_1})\;,
 \nonumber\\
 {\mathcal F} 
 & = &  
 {\mathcal F}_0 + {\mathcal U} \sum\limits_{j=1}^{n} x_j m_j^2\;.
\eq
In general, ${\mathcal U}$ is a positive semi-definite function. 
Its vanishing is related to the  UV sub-divergences of the graph. 
Overall UV divergences, if present,
will always be contained in the  prefactor $\Gamma(\nu-l D/2)$. 
In the Euclidean region, ${\mathcal F}$ is also a positive semi-definite function 
of the Feynman parameters $x_j$.  

As an example we consider the graph in fig.~\ref{example_graph}.
For simplicity we assume that all internal propagators are massless. Then the functions ${\mathcal U}$ and ${\mathcal F}$ read:
\bq
 {\mathcal U} & = & x_{15} x_{23} + x_{15} x_{46} + x_{23} x_{46},
 \nonumber \\
 {\mathcal F} & = & 
  \left( x_1 x_3 x_4 + x_5 x_2 x_6 + x_1 x_5 x_{2346} \right) \left( -p_1^2 \right) 
 \nonumber \\
 & &
  + \left( x_6 x_3 x_5 + x_4 x_1 x_2 + x_4 x_6 x_{1235} \right) \left( -p_2^2 \right) 
 \nonumber \\
 & &
  + \left( x_2 x_4 x_5 + x_3 x_1 x_6 + x_2 x_3 x_{1456} \right) \left( -p_3^2 \right).
\eq
Here we used the notation that $x_{ij...r} = x_i + x_j + ... + x_r$.

Finally let us remark, that in eq.~(\ref{eq0}) we restricted ourselves to scalar integrals, where
the numerator of the integrand is independent of the loop momentum.
A priori more complicated cases, where the loop momentum appears in the numerator might occur.
However, there is a general reduction algorithm, which reduces these tensor integrals
to scalar integrals \cite{Tarasov:1996br,Tarasov:1997kx}.
The price we have to pay is that these scalar integrals involve higher powers of the propagators
and/or have shifted dimensions.
Therefore we considered in eq.~(\ref{shifted_dim}) shifted dimensions
and in eq.~(\ref{eq0}) arbitrary powers of the propagators.
In conclusion, the integrals of the form as in eq.~(\ref{eq0}) 
are the most general loop integrals we have to solve.

% ----------------------------------------------
\section{The Mellin-Barnes transformation}
\label{sect:mellinbarnes}

In sect.~\ref{sect:feynman} we saw that the Feynman parameter integrals 
depend on two graph polynomials ${\mathcal U}$ and ${\mathcal F}$, which are homogeneous functions of the 
Feynman parameters.
In this section we will continue the discussion how these integrals can be performed and exchanged 
against a (multiple) sum over residues.
The case, where the two polynomials are absent is particular simple:
\bq
\label{multi_beta_fct}
 \int\limits_{0}^{1} \left( \prod\limits_{j=1}^{n}\,dx_j\,x_j^{\nu_j-1} \right)
 \delta(1-\sum_{i=1}^n x_i)
 & = & 
 \frac{\prod\limits_{j=1}^{n}\Gamma(\nu_j)}{\Gamma(\nu_1+...+\nu_n)}.
\eq
With the help of 
the Mellin-Barnes transformation we now reduce the general case to eq.~(\ref{multi_beta_fct}).
The Mellin-Barnes transformation reads
\bq
\label{multi_mellin_barnes}
\lefteqn{
\left(A_1 + A_2 + ... + A_n \right)^{-c} 
 = 
 \frac{1}{\Gamma(c)} \frac{1}{\left(2\pi i\right)^{n-1}} 
 \int\limits_{-i\infty}^{i\infty} d\sigma_1 ... \int\limits_{-i\infty}^{i\infty} d\sigma_{n-1}
 } & & \\
 & & 
 \times 
 \Gamma(-\sigma_1) ... \Gamma(-\sigma_{n-1}) \Gamma(\sigma_1+...+\sigma_{n-1}+c)
 \; 
 A_1^{\sigma_1} ...  A_{n-1}^{\sigma_{n-1}} A_n^{-\sigma_1-...-\sigma_{n-1}-c}.
 \nonumber 
\eq
Each contour is such that the poles of $\Gamma(-\sigma)$ are to the right and the poles
of $\Gamma(\sigma+c)$ are to the left.
This transformation can be used to convert the sum of monomials of the polynomials ${\mathcal U}$ and ${\mathcal F}$ into
a product, such that all Feynman parameter integrals are of the form of eq.~(\ref{multi_beta_fct}).
As this transformation converts sums into products it is 
the ``inverse'' of Feynman parametrisation.
Eq.~(\ref{multi_mellin_barnes}) is derived from the theory of Mellin transformations:
Let $h(x)$ be a function which is bounded by a power law for $x\rightarrow 0$ and $x \rightarrow \infty$,
e.g.
\bq
\left| h(x) \right| \le K x^{-c_0} & & \mbox{for}\;\; x \rightarrow 0,
 \nonumber \\
\left| h(x) \right| \le K' x^{c_1} & & \mbox{for}\;\; x \rightarrow \infty.
\eq
Then the Mellin transform is defined for
$c_0 < \mbox{Re}\; \sigma < c_1$
by
\bq
h_{\cal M}(\sigma) & = &
 \int\limits_{0}^\infty dx \; h(x) \; x^{\sigma-1}.
\eq
The inverse Mellin transform is given by
\bq
\label{inversemellin}
h(x) & = & \frac{1}{2\pi i} \int\limits_{\gamma-i\infty}^{\gamma+i\infty}
 d\sigma \; h_{\cal M}(\sigma) \; x^{-\sigma}.
\eq
The integration contour is parallel to the imaginary axis and $c_0 < \mbox{Re}\; \gamma < c_1$.
As an example for the Mellin transform we consider the function 
\bq
h(x) & = & \frac{x^c}{(1+x)^c}
\eq
with Mellin transform $h_{\cal M}(\sigma)=\Gamma(-\sigma) \Gamma(\sigma+c) / \Gamma(c)$.
For $\mbox{Re}(-c) < \mbox{Re} \; \gamma < 0$ we have
\bq
\label{baseMellin}
\frac{x^c}{(1+x)^c}
 & = & 
\frac{1}{2\pi i} \int\limits_{\gamma-i\infty}^{\gamma+i\infty}
 d\sigma \; \frac{\Gamma(-\sigma) \Gamma(\sigma+c)}{\Gamma(c)} \; x^{-\sigma}.
\eq
From eq. (\ref{baseMellin}) one obtains with $x=B/A$ the Mellin-Barnes formula
\bq
\label{simple_mellin_barnes}
\left(A+B\right)^{-c}
 & = & 
\frac{1}{2\pi i} \int\limits_{\gamma-i\infty}^{\gamma+i\infty}
 d\sigma \; \frac{\Gamma(-\sigma) \Gamma(\sigma+c)}{\Gamma(c)} \; A^\sigma B^{-\sigma-c}.
\eq
Eq.~(\ref{multi_mellin_barnes}) is then obtained by repeated use of eq.~(\ref{simple_mellin_barnes}).

With the help of eq.~(\ref{multi_beta_fct}) and eq.~(\ref{multi_mellin_barnes})
we may exchange the Feynman parameter integrals against multiple contour integrals.
A single contour integral is of the form
\bq
\label{MellinBarnesInt}
I
 & = & 
\frac{1}{2\pi i} \int\limits_{\gamma-i\infty}^{\gamma+i\infty}
 d\sigma \; 
 \frac{\Gamma(\sigma+a_1) ... \Gamma(\sigma+a_m)}
      {\Gamma(\sigma+c_2) ... \Gamma(\sigma+c_p)}
 \frac{\Gamma(-\sigma+b_1) ... \Gamma(-\sigma+b_n)}
      {\Gamma(-\sigma+d_1) ... \Gamma(-\sigma+d_q)} 
 \; x^{-\sigma}.
\eq
If $\;\mbox{max}\left( \mbox{Re}(-a_1), ..., \mbox{Re}(-a_m) \right) < \mbox{min}\left( \mbox{Re}(b_1), ..., \mbox{Re}(b_n) \right)$ the contour can be chosen
as a straight line parallel to the imaginary axis with
\bq
\mbox{max}\left( \mbox{Re}(-a_1), ..., \mbox{Re}(-a_m) \right) 
 \;\;\; < \;\;\; \mbox{Re} \; \gamma \;\;\; < \;\;\;
\mbox{min}\left( \mbox{Re}(b_1), ..., \mbox{Re}(b_n) \right),
\eq
otherwise the contour is indented, such that the residues of
$\Gamma(\sigma+a_1)$, ..., $\Gamma(\sigma+a_m)$ are to the right of the contour,
whereas the residues of 
$\Gamma(-\sigma+b_1)$,  ..., $\Gamma(-\sigma+b_n)$ are to the left of the contour.
We further set
\bq
\alpha & = & m+n-p-q,
\nonumber \\
\beta & = & m-n-p+q, 
\nonumber \\
\lambda & = & \mbox{Re} \left( \sum\limits_{j=1}^m a_j
                              +\sum\limits_{j=1}^n b_j
                              -\sum\limits_{j=1}^p c_j
                              -\sum\limits_{j=1}^q d_j \right)
              - \frac{1}{2} \left( m+n-p-q \right).
\eq
Then the integral eq. (\ref{MellinBarnesInt})
converges absolutely for $\alpha >0$ \cite{Erdelyi} and defines an analytic function in
\bq
\left| \mbox{arg} \; x \right| & < & \mbox{min}\left( \pi, \alpha \frac{\pi}{2} \right).
\eq
The integral eq. (\ref{MellinBarnesInt}) is most conveniently evaluated with 
the help of the residuum theorem by closing the contour to the left or to the right.
Therefore we need to know under which conditions the semi-circle at infinity used to close the contour gives a vanishing contribution.
This is obviously the case for $|x|<1$ if we close the contour to the left,
and for $|x|>1$, if we close the contour to the right.
The case $|x|=1$ deserves some special attention. One can show that
in the case $\beta=0$ the semi-circle gives a vanishing contribution, provided
\bq
\lambda & < & -1.
\eq
To sum up all residues which lie inside the contour
it is useful to know the residues of the Gamma function:
\bq
\mbox{res} \; \left( \Gamma(\sigma+a), \sigma=-a-n \right) = \frac{(-1)^n}{n!}, 
 & &
\mbox{res} \; \left( \Gamma(-\sigma+a), \sigma=a+n \right) = -\frac{(-1)^n}{n!}. 
 \nonumber \\
\eq
In general, one obtains (multiple) sum over residues. 
In particular simple cases the contour integrals can be performed in closed form with
the help of two lemmas of Barnes.
Barnes first lemma states that
\bq
\frac{1}{2\pi i} \int\limits_{-i\infty}^{i\infty} d\sigma \;
\Gamma(a+\sigma) \Gamma(b+\sigma) \Gamma(c-\sigma) \Gamma(d-\sigma) 
 =  
\frac{\Gamma(a+c) \Gamma(a+d) \Gamma(b+c) \Gamma(b+d)}{\Gamma(a+b+c+d)},
\;\;\;\;\;\;\;
\hspace*{-15mm}
\nonumber \\
\eq
if none of the poles of $\Gamma(a+\sigma) \Gamma(b+\sigma)$ coincides with the
ones from $\Gamma(c-\sigma) \Gamma(d-\sigma)$.
Barnes second lemma reads
\bq
\lefteqn{
\frac{1}{2\pi i} \int\limits_{-i\infty}^{i\infty} d\sigma \;
\frac{\Gamma(a+\sigma) \Gamma(b+\sigma) \Gamma(c+\sigma) \Gamma(d-\sigma) \Gamma(e-\sigma)}
{\Gamma(a+b+c+d+e+\sigma)} } & & \nonumber \\
& = & 
\frac{\Gamma(a+d) \Gamma(b+d) \Gamma(c+d) 
      \Gamma(a+e) \Gamma(b+e) \Gamma(c+e)}
{\Gamma(a+b+d+e) \Gamma(a+c+d+e) \Gamma(b+c+d+e)}.
\eq
Although the Mellin-Barnes transformation has been known for a long time, 
the method has seen a revival in applications in recent 
years \cite{Boos:1990rg,Davydychev:1990jt,Davydychev:1990cq,Smirnov:1999gc,Smirnov:1999wz,Tausk:1999vh,Smirnov:2000vy,Smirnov:2000ie,Smirnov:2003vi,Bierenbaum:2003ud,Heinrich:2004iq,Friot:2005cu,Bern:2005iz,Anastasiou:2005cb,Czakon:2005rk,Gluza:2007rt}.

% ----------------------------------------------
\section{Shuffle algebras}
\label{sect:shuffle}

Before we continue the discussion of loop integrals, it is useful to discuss first
shuffle algebras and generalisations thereof from an algebraic viewpoint.
Consider a set of letters $A$. The set $A$ is called the alphabet.
A word is an ordered sequence of letters:
\bq
 w & = & l_1 l_2 ... l_k.
\eq
The word of length zero is denoted by $e$.
Let $K$ be a field and consider the vector space of words over $K$.
A shuffle algebra ${\cal A}$ on the vector space of words is defined by
\bq
\left( l_1 l_2 ... l_k \right) \cdot 
 \left( l_{k+1} ... l_r \right) & = &
 \sum\limits_{\mbox{\tiny shuffles} \; \sigma} l_{\sigma(1)} l_{\sigma(2)} ... l_{\sigma(r)},
\eq
where the sum runs over all permutations $\sigma$, which preserve the relative order
of $1,2,...,k$ and of $k+1,...,r$.
The name ``shuffle algebra'' is related to the analogy of shuffling cards: If a deck of cards
is split into two parts and then shuffled, the relative order within the two individual parts
is conserved.
The empty word $e$ is the unit in this algebra:
\bq
 e \cdot w = w \cdot e = w.
\eq
A recursive definition of the shuffle product is given by
\bq
\label{def_recursive_shuffle}
\left( l_1 l_2 ... l_k \right) \cdot \left( l_{k+1} ... l_r \right) & = &
 l_1 \left[ \left( l_2 ... l_k \right) \cdot \left( l_{k+1} ... l_r \right) \right]
+
 l_{k+1} \left[ \left( l_1 l_2 ... l_k \right) \cdot \left( l_{k+2} ... l_r \right) \right]
\eq
It is well known fact that the shuffle algebra is actually a (non-cocommutative) Hopf algebra \cite{Reutenauer}.
In this context let us briefly review the definitions of a coalgebra, a bialgebra and a Hopf algebra,
which are closely related:
First note that the unit in an algebra can be viewed as a map from $K$ to $A$ and that the multiplication
can be viewed as a map from the tensor product $A \otimes A$ to $A$ (e.g. one takes two elements
from $A$, multiplies them and gets one element out). 

A coalgebra has instead of multiplication and unit the dual structures:
a comultiplication $\Delta$ and a counit $\bar{e}$.
The counit is a map from $A$ to $K$, whereas comultiplication is a map from $A$ to
$A \otimes A$.
Note that comultiplication and counit go in the reverse direction compared to multiplication
and unit.
We will always assume that the comultiplication is coassociative.
The general form of the coproduct is
\bq
\Delta(a) & = & \sum\limits_i a_i^{(1)} \otimes a_i^{(2)},
\eq
where $a_i^{(1)}$ denotes an element of $A$ appearing in the first slot of $A \otimes A$ and
$a_i^{(2)}$ correspondingly denotes an element of $A$ appearing in the second slot.
Sweedler's notation \cite{Sweedler} consists in dropping the dummy index $i$ and the summation symbol:
\bq
\Delta(a) & = & 
a^{(1)} \otimes a^{(2)}
\eq 
The sum is implicitly understood. This is similar to Einstein's summation convention, except
that the dummy summation index $i$ is also dropped. The superscripts ${}^{(1)}$ and ${}^{(2)}$ 
indicate that a sum is involved.

A bialgebra is an algebra and a coalgebra at the same time,
such that the two structures are compatible with each other.
Using Sweedler's notation,
the compatibility between the multiplication and comultiplication is express\-ed as
\bq
\label{bialg}
 \Delta\left( a \cdot b \right)
 & = &
\left( a^{(1)} \cdot b^{(1)} \right)
 \otimes \left( a^{(2)} \cdot b^{(2)} \right).
\eq

A Hopf algebra is a bialgebra with an additional map from $A$ to $A$, called the 
antipode ${\cal S}$, which fulfils
\bq
a^{(1)} \cdot {\cal S}\left( a^{(2)} \right)
=
{\cal S}\left(a^{(1)}\right) \cdot a^{(2)} 
= 0 & &\;\;\; \mbox{for} \; a \neq e.
\eq

With this background at hand we can now state the coproduct, the counit and the antipode for the
shuffle algebra:
The counit $\bar{e}$ is given by:
\bq
\bar{e}\left( e\right) = 1, \;\;\;
& &
\bar{e}\left( l_1 l_2 ... l_n\right) = 0.
\eq
The coproduct $\Delta$ is given by:
\bq
\Delta\left( l_1 l_2 ... l_k \right) 
& = & \sum\limits_{j=0}^k \left( l_{j+1} ... l_k \right) \otimes \left( l_1 ... l_j \right).
\eq
The antipode ${\cal S}$ is given by:
\bq
{\cal S}\left( l_1 l_2 ... l_k \right) & = & (-1)^k \; l_k l_{k-1} ... l_2 l_1.
\eq
The shuffle algebra is generated by the Lyndon words.
If one introduces a lexicographic ordering on the letters of the alphabet
$A$, a Lyndon word is defined by the property
\bq
w < v
\eq
for any sub-words $u$ and $v$ such that $w= u v$.

\begin{figure}
\begin{center}
\begin{picture}(300,65)(0,0)
\put(10,10){\vector(1,0){50}}
\put(10,10){\vector(0,1){50}}
\Text(60,5)[t]{$t_1$}
\Text(5,60)[r]{$t_2$}
\Line(10,50)(50,50)
\Line(50,50)(50,10)
\Line(10,20)(20,10)
\Line(10,30)(30,10)
\Line(10,40)(40,10)
\Line(10,50)(50,10)
\Line(20,50)(50,20)
\Line(30,50)(50,30)
\Line(40,50)(50,40)
\Text(80,30)[c]{$=$}
\put(110,10){\vector(1,0){50}}
\put(110,10){\vector(0,1){50}}
\Text(160,5)[t]{$t_1$}
\Text(105,60)[r]{$t_2$}
\Line(110,10)(150,50)
\Line(150,50)(150,10)
\Line(120,10)(120,20)
\Line(130,10)(130,30)
\Line(140,10)(140,40)
\Text(180,30)[c]{$+$}
\put(210,10){\vector(1,0){50}}
\put(210,10){\vector(0,1){50}}
\Text(260,5)[t]{$t_1$}
\Text(205,60)[r]{$t_2$}
\Line(210,50)(250,50)
\Line(250,50)(210,10)
\Line(210,20)(220,20)
\Line(210,30)(230,30)
\Line(210,40)(240,40)
\end{picture}
\caption{\label{proof_shuffle} Sketch of the proof for the shuffle product of two iterated integrals.
The integral over the square is replaced by two
integrals over the upper and lower triangle.}
\end{center}
\end{figure}
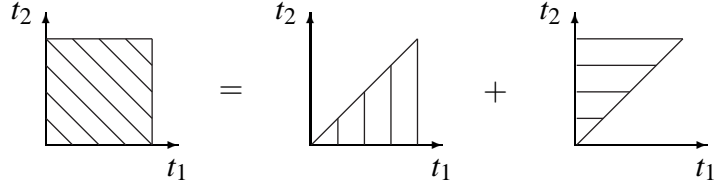

An important example for a shuffle algebra are iterated integrals.
Let $[a, b]$ be a segment of the real line and $f_1$, $f_2$, ... functions
on this interval.
Let us define the following iterated integrals:
\bq
 I(f_1,f_2,...,f_k;a,b) 
 & = &
 \int\limits_a^b f_1(t_1) dt_1 \int\limits_a^{t_1} f_2(t_2) dt_2
 ...
 \int\limits_a^{t_{k-1}} f_k(t_k) dt_k
\eq
For fixed $a$ and $b$ we have a shuffle algebra:
\bq
 I(f_1,f_2,...,f_k;a,b) \cdot I(f_{k+1},...,f_r; a,b) & = &
 \sum\limits_{\mbox{\tiny shuffles} \; \sigma} I(f_{\sigma(1)},f_{\sigma(2)},...,f_{\sigma(r)};a,b),
\eq
where the sum runs over all permutations $\sigma$, which preserve the relative order
of $1,2,...,k$ and of $k+1,...,r$.
The proof is sketched in fig.~\ref{proof_shuffle}.
The two outermost integrations are recursively replaced by integrations over the upper and lower triangle.

We now consider generalisations of shuffle algebras. Assume that for the set of letters we have an additional
operation
\bq
 (.,.) & : & A \otimes A \rightarrow A,
 \nonumber \\
       & &  l_1 \otimes l_2 \rightarrow (l_1, l_2),
\eq
which is commutative and associative.
Then we can define a new product of words recursively through
\bq
\label{def_recursive_quasi_shuffle}
\left( l_1 l_2 ... l_k \right) \ast \left( l_{k+1} ... l_r \right) & = &
 l_1 \left[ \left( l_2 ... l_k \right) \ast \left( l_{k+1} ... l_r \right) \right]
+
 l_{k+1} \left[ \left( l_1 l_2 ... l_k \right) \ast \left( l_{k+2} ... l_r \right) \right]
 \nonumber \\
 & &
+
(l_1,l_{k+1}) \left[ \left( l_2 ... l_k \right) \ast \left( l_{k+2} ... l_r \right) \right]
\eq
This product is a generalisation of the shuffle product and differs from the recursive
definition of the shuffle product in eq.~(\ref{def_recursive_shuffle}) through the extra term in the last line.
This modified product is known under the names quasi-shuffle product \cite{Hoffman},
mixable shuffle product \cite{Guo}
or stuffle product \cite{Borwein}.
Quasi-shuffle algebras are Hopf algebras.
Comultiplication and counit are defined as for the shuffle algebras.
The counit $\bar{e}$ is given by:
\bq
\bar{e}\left( e\right) = 1, \;\;\;
& &
\bar{e}\left( l_1 l_2 ... l_n\right) = 0.
\eq
The coproduct $\Delta$ is given by:
\bq
\Delta\left( l_1 l_2 ... l_k \right) 
& = & \sum\limits_{j=0}^k \left( l_{j+1} ... l_k \right) \otimes \left( l_1 ... l_j \right).
\eq
The antipode ${\cal S}$ is recursively defined through
\bq
{\cal S}\left( l_1 l_2 ... l_k \right) & = & 
 - l_1 l_2 ... l_k
 - \sum\limits_{j=1}^{k-1} {\cal S}\left( l_{j+1} ... l_k \right) \ast \left( l_1 ... l_j \right).
\eq
An example for a quasi-shuffle algebra are nested sums.
Let $n_a$ and $n_b$ be integers with $n_a<n_b$ and let $f_1$, $f_2$, ... be functions
defined on the integers.
We consider the following nested sums:
\bq
 S(f_1,f_2,...,f_k;n_a,n_b) 
 & = &
 \sum\limits_{i_1=n_a}^{n_b} f_1(i_1) \sum\limits_{i_2=n_a}^{i_1-1} f_2(i_2) 
 ...
 \sum\limits_{i_k=n_a}^{i_{k-1}-1} f_k(i_k)
\eq
For fixed $n_a$ and $n_b$ we have a quasi-shuffle algebra:
\bq
\label{quasi_shuffle_multiplication}
\lefteqn{
 S(f_1,f_2,...,f_k;n_a,n_b) \ast S(f_{k+1},...,f_r; n_a,n_b) 
= } & &
 \nonumber \\
 & &
   \sum\limits_{i_1=n_a}^{n_b} f_1(i_1) \; S(f_2,...,f_k;n_a,i_1-1) \ast S(f_{k+1},...,f_r; n_a,i_1-1)
 \nonumber \\
 & &
 +  \sum\limits_{j_1=n_a}^{n_b} f_k(j_1) \; S(f_1,f_2,...,f_k;n_a,j_1-1) \ast S(f_{k+2},...,f_r; n_a,j_1-1)
 \nonumber \\
 & &
 +  \sum\limits_{i=n_a}^{n_b} f_1(i) f_k(i) \; S(f_2,...,f_k;n_a,i-1) \ast S(f_{k+2},...,f_r; n_a,i-1)
\eq
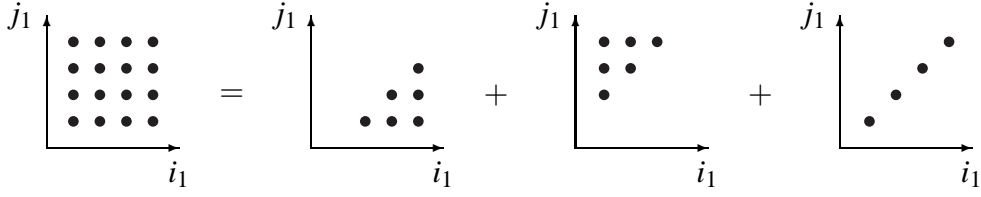
\begin{figure}
\begin{center}
\begin{picture}(400,65)(0,0)
\put(10,10){\vector(1,0){50}}
\put(10,10){\vector(0,1){50}}
\Text(60,5)[t]{$i_1$}
\Text(5,60)[r]{$j_1$}
\Vertex(20,20){2}
\Vertex(30,20){2}
\Vertex(40,20){2}
\Vertex(50,20){2}
\Vertex(20,30){2}
\Vertex(30,30){2}
\Vertex(40,30){2}
\Vertex(50,30){2}
\Vertex(20,40){2}
\Vertex(30,40){2}
\Vertex(40,40){2}
\Vertex(50,40){2}
\Vertex(20,50){2}
\Vertex(30,50){2}
\Vertex(40,50){2}
\Vertex(50,50){2}
\Text(80,30)[c]{$=$}
\put(110,10){\vector(1,0){50}}
\put(110,10){\vector(0,1){50}}
\Text(160,5)[t]{$i_1$}
\Text(105,60)[r]{$j_1$}
\Vertex(130,20){2}
\Vertex(140,20){2}
\Vertex(150,20){2}
\Vertex(140,30){2}
\Vertex(150,30){2}
\Vertex(150,40){2}
\Text(180,30)[c]{$+$}
\put(210,10){\vector(1,0){50}}
\put(210,10){\vector(0,1){50}}
\Text(260,5)[t]{$i_1$}
\Text(205,60)[r]{$j_1$}
\Vertex(220,30){2}
\Vertex(220,40){2}
\Vertex(230,40){2}
\Vertex(220,50){2}
\Vertex(230,50){2}
\Vertex(240,50){2}
\Text(280,30)[c]{$+$}
\put(310,10){\vector(1,0){50}}
\put(310,10){\vector(0,1){50}}
\Text(360,5)[t]{$i_1$}
\Text(305,60)[r]{$j_1$}
\Vertex(320,20){2}
\Vertex(330,30){2}
\Vertex(340,40){2}
\Vertex(350,50){2}
\end{picture}
\caption{\label{proof} Sketch of the proof for the quasi-shuffle product of nested sums. 
The sum over the square is replaced by
the sum over the three regions on the r.h.s.}
\end{center}
\end{figure}
Note that the product of two letters corresponds to the point-wise product of the two functions:
\bq
 ( f_i, f_j ) \; (n) & = & f_i(n) f_j(n).
\eq
The proof that nested sums obey the quasi-shuffle algebra is sketched in Fig. \ref{proof}.
The outermost sums of the nested sums on the l.h.s of (\ref{quasi_shuffle_multiplication}) are split into the three
regions indicated in Fig. \ref{proof}.

% ----------------------------------------------
\section{Multiple polylogarithms}
\label{sect:polylog}

In the previous section we have seen that iterated integrals form a shuffle algebra, while
nested sums form a quasi-shuffle algebra.
In this context multiple polylogarithms form an interesting class of functions.
They have a representation as iterated integrals as well as nested sums.
Therefore multiple polylogarithms form a shuffle algebra as well as a quasi-shuffle algebra.
The two algebra structures are independent.
Let us start with the representation as nested sums.
The multiple polylogarithms are defined by
\bq 
\label{multipolylog2}
 \mbox{Li}_{m_1,...,m_k}(x_1,...,x_k)
  & = & \sum\limits_{i_1>i_2>\ldots>i_k>0}
     \frac{x_1^{i_1}}{{i_1}^{m_1}}\ldots \frac{x_k^{i_k}}{{i_k}^{m_k}}.
\eq
The multiple polylogarithms are generalisations of
the classical polylogarithms 
$
\mbox{Li}_n(x)
$ 
\cite{lewin:book},
whose most prominent examples are
\bq
 \mbox{Li}_1(x) = \sum\limits_{i_1=1}^\infty \frac{x^{i_1}}{i_1} = -\ln(1-x),
 & &
 \mbox{Li}_2(x) = \sum\limits_{i_1=1}^\infty \frac{x^{i_1}}{i_1^2},
\eq 
as well as
Nielsen's generalised polylogarithms \cite{Nielsen}
\bq
S_{n,p}(x) & = & \mbox{Li}_{n+1,1,...,1}(x,\underbrace{1,...,1}_{p-1}),
\eq
and the harmonic polylogarithms \cite{Remiddi:1999ew}
\bq
\label{harmpolylog}
H_{m_1,...,m_k}(x) & = & \mbox{Li}_{m_1,...,m_k}(x,\underbrace{1,...,1}_{k-1}).
\eq
Multiple polylogarithms have been studied extensively in the literature
by physicists
\cite{Remiddi:1999ew,Vermaseren:1998uu,Gehrmann:2000zt,Gehrmann:2001pz,Gehrmann:2001jv,Gehrmann:2002zr,Moch:2001zr,Blumlein:1998if,Blumlein:2003gb,Weinzierl:2004bn,Vollinga:2004sn,Korner:2005qz,Kalmykov:2006hu,Maitre:2007kp}
and mathematicians
\cite{Borwein,Hain,Goncharov,Goncharov:2001,Goncharov:2002,Goncharov:2002b,Gangl:2000,Gangl:2002,Minh:2000,Cartier:2001,Ecalle,Racinet:2002}.

In addition, multiple polylogarithms have an integral representation. 
To discuss the integral representation it is convenient to 
introduce for $z_k \neq 0$
the following functions
\bq
\label{Gfuncdef}
G(z_1,...,z_k;y) & = &
 \int\limits_0^y \frac{dt_1}{t_1-z_1}
 \int\limits_0^{t_1} \frac{dt_2}{t_2-z_2} ...
 \int\limits_0^{t_{k-1}} \frac{dt_k}{t_k-z_k}.
\eq
In this definition 
one variable is redundant due to the following scaling relation:
\bq
G(z_1,...,z_k;y) & = & G(x z_1, ..., x z_k; x y)
\eq
If one further defines
\bq
g(z;y) & = & \frac{1}{y-z},
\eq
then one has
\bq
\frac{d}{dy} G(z_1,...,z_k;y) & = & g(z_1;y) G(z_2,...,z_k;y)
\eq
and
\bq
\label{Grecursive}
G(z_1,z_2,...,z_k;y) & = & \int\limits_0^y dt \; g(z_1;t) G(z_2,...,z_k;t).
\eq
One can slightly enlarge the set and define
$G(0,...,0;y)$ with $k$ zeros for $z_1$ to $z_k$ to be
\bq
\label{trailingzeros}
G(0,...,0;y) & = & \frac{1}{k!} \left( \ln y \right)^k.
\eq
This permits us to allow trailing zeros in the sequence
$(z_1,...,z_k)$ by defining the function $G$ with trailing zeros via (\ref{Grecursive}) 
and (\ref{trailingzeros}).
To relate the multiple polylogarithms to the functions $G$ it is convenient to introduce
the following short-hand notation:
\bq
\label{Gshorthand}
G_{m_1,...,m_k}(z_1,...,z_k;y)
 & = &
 G(\underbrace{0,...,0}_{m_1-1},z_1,...,z_{k-1},\underbrace{0...,0}_{m_k-1},z_k;y)
\eq
Here, all $z_j$ for $j=1,...,k$ are assumed to be non-zero.
One then finds
\bq
\label{Gintrepdef}
\mbox{Li}_{m_1,...,m_k}(x_1,...,x_k)
& = & (-1)^k 
 G_{m_1,...,m_k}\left( \frac{1}{x_1}, \frac{1}{x_1 x_2}, ..., \frac{1}{x_1...x_k};1 \right).
\eq
The inverse formula reads
\bq
G_{m_1,...,m_k}(z_1,...,z_k;y) & = & 
 (-1)^k \; \mbox{Li}_{m_1,...,m_k}\left(\frac{y}{z_1}, \frac{z_1}{z_2}, ..., \frac{z_{k-1}}{z_k}\right).
\eq
Eq. (\ref{Gintrepdef}) together with 
(\ref{Gshorthand}) and (\ref{Gfuncdef})
defines an integral representation for the multiple polylogarithms.
To make this more explicit I first introduce some notation for iterated integrals
\bq
\int\limits_0^\Lambda \frac{dt}{t-a_n} \circ ... \circ \frac{dt}{t-a_1} & = & 
\int\limits_0^\Lambda \frac{dt_n}{t_n-a_n} \int\limits_0^{t_n} \frac{dt_{n-1}}{t_{n-1}-a_{n-1}} \times ... \times \int\limits_0^{t_2} \frac{dt_1}{t_1-a_1}
\;\;\;\;\;\;\;\;
\eq
and the short hand notation:
\bq
\int\limits_0^\Lambda \left( \frac{dt}{t} \circ \right)^{m} \frac{dt}{t-a}
& = & 
\int\limits_0^\Lambda 
\underbrace{\frac{dt}{t} \circ ... \frac{dt}{t}}_{m \;\mbox{times}} \circ \frac{dt}{t-a}.
\eq
The integral representation for $\mbox{Li}_{m_1,...,m_k}(x_1,...,x_k)$ reads then
\bq
\label{intrepII}
\lefteqn{
\mbox{Li}_{m_1,...,m_k}(x_1,...,x_k) = 
 (-1)^k \int\limits_0^1 \left( \frac{dt}{t} \circ \right)^{m_1-1} \frac{dt}{t-b_1} 
 } \nonumber \\
 & & 
 \circ \left( \frac{dt}{t} \circ \right)^{m_2-1} \frac{dt}{t-b_2}
 \circ ... \circ
 \left( \frac{dt}{t} \circ \right)^{m_k-1} \frac{dt}{t-b_k},
\eq
where the $b_j$'s are related to the $x_j$'s 
\bq
b_j & = & \frac{1}{x_1 x_2 ... x_j}.
\eq

Up to now we treated multiple polylogarithms from an algebraic point of view.
Equally important are the analytical properties, which are needed for an efficient numerical 
evaluation.
As an example I first discuss the numerical evaluation of the dilogarithm \cite{'tHooft:1979xw}:
\bq
\mbox{Li}_{2}(x) & = & - \int\limits_{0}^{x} dt \frac{\ln(1-t)}{t}
 = \sum\limits_{n=1}^{\infty} \frac{x^{n}}{n^{2}}
\eq
The power series expansion can be evaluated numerically, provided $|x| < 1.$
Using the functional equations 
\bq
\mbox{Li}_2(x) & = & -\mbox{Li}_2\left(\frac{1}{x}\right) -\frac{\pi^2}{6} -\frac{1}{2} \left( \ln(-x) \right)^2,
 \nonumber \\
\mbox{Li}_2(x) & = & -\mbox{Li}_2(1-x) + \frac{\pi^2}{6} -\ln(x) \ln(1-x).
\eq
any argument of the dilogarithm can be mapped into the region
$|x| \le 1$ and
$-1 \leq \mbox{Re}(x) \leq 1/2$.
The numerical computation can be accelerated  by using an expansion in $[-\ln(1-x)]$ and the
Bernoulli numbers $B_i$:
\bq
\mbox{Li}_2(x) & = & \sum\limits_{i=0}^\infty \frac{B_i}{(i+1)!} \left( - \ln(1-x) \right)^{i+1}.
\eq
The generalisation to multiple polylogarithms proceeds along the same lines \cite{Vollinga:2004sn}:
Using the integral representation
\bq
\lefteqn{
G_{m_1,...,m_k}\left(z_1,z_2,...,z_k;y\right)
 = } & &
 \\
 & &
 \int\limits_0^y \left( \frac{dt}{t} \circ \right)^{m_1-1} \frac{dt}{t-z_1}
 \left( \frac{dt}{t} \circ \right)^{m_2-1} \frac{dt}{t-z_2}
 ...
 \left( \frac{dt}{t} \circ \right)^{m_k-1} \frac{dt}{t-z_k}
 \nonumber 
\eq
one
transforms all arguments into a region, where one has a converging power series expansion:
\bq
G_{m_1,...,m_k}\left(z_1,...,z_k;y\right) 
 & = &
 \sum\limits_{j_1=1}^\infty
 ... 
 \sum\limits_{j_k=1}^\infty 
 \frac{1}{\left(j_1+...+j_k\right)^{m_1}} \left( \frac{y}{z_1} \right)^{j_1}
 \nonumber \\
 & & 
 \times 
 \frac{1}{\left(j_2+...+j_k\right)^{m_2}} \left( \frac{y}{z_2} \right)^{j_2}
 ...
 \frac{1}{\left(j_k\right)^{m_k}} \left( \frac{y}{z_k} \right)^{j_k}.
\;\;\;\;\;\;
\eq 
The multiple polylogarithms satisfy the H\"older convolution \cite{Borwein}.
For $z_1 \neq 1$ and $z_w \neq 0$ this identity reads
\bq
\label{defhoelder}
\lefteqn{
G\left(z_1,...,z_w; 1 \right) 
 = } & & 
 \\
 & &
 \sum\limits_{j=0}^w \left(-1\right)^j 
  G\left(1-z_j, 1-z_{j-1},...,1-z_1; 1 - \frac{1}{p} \right)
  G\left( z_{j+1},..., z_w; \frac{1}{p} \right).
 \nonumber 
\eq
The H\"older convolution can be used to accelerate the 
convergence for the series
representation of the multiple polylogarithms.

% ----------------------------------------------
\section{Laurent expansion of Feynman integrals}
\label{sect:laurent}

Let us return to the question on how to compute Feynman integrals.
In section \ref{sect:mellinbarnes} we saw how to obtain from the Mellin-Barnes
transformation (multiple) sums by closing the integration contours and summing
up the residues.
As a simple example let us consider that the sum of residues is equal to
\bq
\label{examplehypergeom}
\sum\limits_{i=0}^\infty
 \frac{\Gamma(i+a_1+t_1\eps)\Gamma(i+a_2+t_2\eps)}{\Gamma(i+1)\Gamma(i+a_3+t_3\eps)}
 x^i
\eq
Here $a_1$, $a_2$ and $a_3$ are assumed to be integers.
Up to prefactors the expression in eq.~(\ref{examplehypergeom}) is a hyper-geometric function ${}_2F_1$.
We are interested in the Laurent expansion of this expression in the small parameter
$\eps$.
The basic formula for the expansion of Gamma functions reads
\bq
\label{expansiongamma}
\lefteqn{
\hspace*{-1cm}
\Gamma(n+\eps)  = \Gamma(1+\eps) \Gamma(n)
 \left[
        1 + \eps Z_1(n-1) + \eps^2 Z_{11}(n-1)
 \right.
} \nonumber \\
 & & \left.
          + \eps^3 Z_{111}(n-1) + ... + \eps^{n-1} Z_{11...1}(n-1)
 \right],
\eq
where $Z_{m_1,...,m_k}(n)$ are Euler-Zagier sums
defined by
\bq
 Z_{m_1,...,m_k}(n) & = &
  \sum\limits_{n \ge i_1>i_2>\ldots>i_k>0}
     \frac{1}{{i_1}^{m_1}}\ldots \frac{1}{{i_k}^{m_k}}.
\eq
This motivates the following definition of a special form of nested sums, called 
$Z$-sums:
\bq 
\label{definition}
  Z(n;m_1,...,m_k;x_1,...,x_k) & = & \sum\limits_{n\ge i_1>i_2>\ldots>i_k>0}
     \frac{x_1^{i_1}}{{i_1}^{m_1}}\ldots \frac{x_k^{i_k}}{{i_k}^{m_k}}.
\eq
$k$ is called the depth of the $Z$-sum and $w=m_1+...+m_k$ is called the weight.
If the sums go to infinity ($n=\infty$) the $Z$-sums are multiple polylogarithms:
\bq
\label{multipolylog}
Z(\infty;m_1,...,m_k;x_1,...,x_k) & = & \mbox{Li}_{m_1,...,m_k}(x_1,...,x_k).
\eq
For $x_1=...=x_k=1$ the definition reduces to the Euler-Zagier sums \cite{Euler,Zagier}:
\bq
Z(n;m_1,...,m_k;1,...,1) & = & Z_{m_1,...,m_k}(n).
\eq
For $n=\infty$ and $x_1=...=x_k=1$ the sum is a multiple $\zeta$-value \cite{Borwein}:
\bq
Z(\infty;m_1,...,m_k;1,...,1) & = & \zeta_{m_1,...,m_k}.
\eq
The usefulness of the $Z$-sums lies in the fact, that they interpolate between
multiple polylogarithms and Euler-Zagier sums.
The $Z$-sums form a quasi-shuffle algebra.

Using $\Gamma(x+1) = x \Gamma(x)$, partial fractioning and an adjustment of the
summation index one can transform eq.~(\ref{examplehypergeom}) into terms of the form
\bq
\sum\limits_{i=1}^\infty
 \frac{\Gamma(i+t_1\eps)\Gamma(i+t_2\eps)}{\Gamma(i)\Gamma(i+t_3\eps)}
 \frac{x^i}{i^m},
\eq
where $m$ is an integer.
Now using eq.~(\ref{expansiongamma})
one obtains
\bq
\Gamma(1+\eps) 
\sum\limits_{i=1}^\infty
 \frac{\left(1+\eps t_1 Z_1(i-1)+...\right) \left(1+\eps t_2 Z_1(i-1)+...\right)}
      {\left(1+\eps t_3 Z_1(i-1)+...\right)}
 \frac{x^i}{i^m}.
\eq
Inverting the power series in the denominator and truncating in $\eps$ one obtains
in each order in $\eps$ terms of the form
\bq
\label{exZ1}
\sum\limits_{i=1}^\infty
 \frac{x^i}{i^m}
 Z_{m_1 ... m_k}(i-1) Z_{m_1' ... m_l'}(i-1) Z_{m_1'' ... m_n''}(i-1)
\eq
Using the quasi-shuffle product for $Z$-sums the three Euler-Zagier sums
can be reduced to single Euler-Zagier sums and one finally arrives at terms of the form
\bq
\label{exZ2}
\sum\limits_{i=1}^\infty
 \frac{x^i}{i^m}
 Z_{m_1 ... m_k}(i-1),
\eq
which are special cases of multiple polylogarithms, called harmonic polylogarithms $H_{m,m_1,...,m_k}(x)$.
This completes the algorithm for the expansion in $\eps$ for sums of the form as in eq.~(\ref{examplehypergeom}).

The Hopf algebra of $Z$-sums has additional structures if we allow expressions
of the form
\bq
\label{augmented}
\frac{x_0^n}{n^{m_0}} Z(n;m_1,...,m_k;x_1,...,x_k),
\eq
e.g. $Z$-sums multiplied by a letter.
Then the following convolution product
\bq
\label{convolution}
 \sum\limits_{i=1}^{n-1} \; \frac{x^i}{i^m} Z(i-1;...)
                         \; \frac{y^{n-i}}{(n-i)^{m'}} Z(n-i-1;...)
\eq
can again be expressed in terms of expressions of the form (\ref{augmented}).
In addition there is a conjugation, e.g. sums of the form 
\bq
\label{conjugation}
 - \sum\limits_{i=1}^n 
       \left( \begin{array}{c} n \\ i \\ \end{array} \right)
       \left( -1 \right)^i
       \; \frac{x^i}{i^m} Z(i;...)
\eq
can also be reduced to terms of the form (\ref{augmented}).
The name conjugation stems from the following fact:
To any function $f(n)$ of an integer variable $n$ one can define
a conjugated function $C \circ f(n)$ as the following sum
\bq
C \circ f(n) & = & \sum\limits_{i=1}^n 
       \left( \begin{array}{c} n \\ i \\ \end{array} \right)
       (-1)^i f(i).
\eq
Then conjugation satisfies the following two properties:
\bq
C \circ 1 & = & 1,
 \nonumber \\
C \circ C \circ f(n) & = & f(n).
\eq
Finally there is the combination of conjugation and convolution,
e.g. sums of the form 
\bq
\label{conjugationconvolution}
 - \sum\limits_{i=1}^{n-1} 
       \left( \begin{array}{c} n \\ i \\ \end{array} \right)
       \left( -1 \right)^i
       \; \frac{x^i}{i^m} Z(i;...)
       \; \frac{y^{n-i}}{(n-i)^{m'}} Z(n-i;...)
\eq
can also be reduced to terms of the form (\ref{augmented}).
These properties can be used to expand more complicated transcendental functions like
\bq
\label{type_B}
     \sum\limits_{i=0}^\infty 
     \sum\limits_{j=0}^\infty 
       \frac{\Gamma(i+a_1)}{\Gamma(i+a_1')} ...
       \frac{\Gamma(i+a_k)}{\Gamma(i+a_k')}
       \frac{\Gamma(j+b_1)}{\Gamma(j+b_1')} ...
       \frac{\Gamma(j+b_l)}{\Gamma(j+b_l')}
       \frac{\Gamma(i+j+c_1)}{\Gamma(i+j+c_1')} ...
       \frac{\Gamma(i+j+c_m)}{\Gamma(i+j+c_m')}
       \; x^i y^j
\eq
or
\bq
\label{type_D}
     \sum\limits_{i=0}^\infty 
     \sum\limits_{j=0}^\infty 
       \left( \begin{array}{c} i+j \\ j \\ \end{array} \right)
       \frac{\Gamma(i+a_1)}{\Gamma(i+a_1')} ...
       \frac{\Gamma(i+a_k)}{\Gamma(i+a_k')}
       \frac{\Gamma(j+b_1)}{\Gamma(j+b_1')} ...
       \frac{\Gamma(j+b_l)}{\Gamma(j+b_l')}
       \frac{\Gamma(i+j+c_1)}{\Gamma(i+j+c_1')} ...
       \frac{\Gamma(i+j+c_m)}{\Gamma(i+j+c_m')}
       \; x^i y^j.
 \nonumber \\
\eq
Examples for functions of this type are the first and second Appell function $F_1$ and $F_2$.
Note that in these examples there are always as many Gamma functions in the numerator
as in the denominator.
We assume that all $a_n$, $a_n'$, $b_n$, $b_n'$, $c_n$  and $c_n'$ are 
of the form ``integer $+ \;\mbox{const} \cdot \eps$''.

The first type can be generalised to the form ``rational number $+ \;\mbox{const} \cdot \eps$'',
if the Gamma functions always occur in ratios of the form
\bq
\label{rational_balanced}
 \frac{\Gamma(n+a-\frac{p}{q} +b \eps)}
      {\Gamma(n+c-\frac{p}{q} +d \eps)},
\eq
where the same rational number $p/q$ occurs in the numerator and in the denominator
\cite{Weinzierl:2004bn}.
In this case we have to replace eq.~(\ref{expansiongamma})
by
\bq
\Gamma\left( n+1-\frac{p}{q}+\eps \right)
 & = &
\frac{\Gamma\left( 1-\frac{p}{q}+\eps\right)\Gamma\left( n+1-\frac{p}{q}\right)}{\Gamma\left( 1-\frac{p}{q} \right)}
 \\
 & & \times
\exp \left( - \frac{1}{q} \sum\limits_{l=0}^{q-1}
            \left( r_q^l \right)^p
            \sum\limits_{k=1}^\infty
             \eps^k \frac{(-q)^k}{k}
             Z( q \cdot n; k; r_q^l )
      \right),
 \nonumber
\eq
which introduces the $q$-th roots of unity
\bq
r_q^p & = & \exp \left( \frac{2 \pi i p}{q} \right).
\eq
In summary these techniques allow a systematic procedure for the computation of Feynman integrals, if certain
conditions are met.
These conditions require that factors of Gamma functions are balanced 
like in eq.~(\ref{type_B}) or eq.~(\ref{type_D}) \cite{Moch:2001zr,Weinzierl:2004bn}.
The algebraic properties of nested sums and iterated integrals discussed here
are well-suited for an
implementation into a computer algebra system
and several packages for these manipulations exist \cite{Vermaseren:1998uu,Weinzierl:2002hv,Moch:2005uc,Maitre:2005uu,Huber:2005yg}.

% ----------------------------------------------
\section{Conclusions}
\label{sect:concl}

In this article I discussed the mathematical structures underlying the computation
of Feynman loop integrals.
One encounters iterated structures as nested sums or iterated integrals, which form
a Hopf algebra with a shuffle or quasi-shuffle product.
Of particular importance are multiple polylogarithms. 
The algebraic properties of these functions are very rich: They form at the same time
a shuffle algebra as well as a quasi-shuffle algebra.
Based on these algebraic structures I discussed algorithms which evaluate Feynman integrals
to multiple polylogarithms.

% ----------------------------------------------
% references

\end{document}